# CommUnet: U-net decoder for convolutional codes in communication

Noam Katz

Tel Aviv University

Noamkatz1@mail.tau.ac.il

*Abstract-* **In recent years, deep neural networks have played a major role solving various challenges in two dimensional image processing. Fully Convolutional Networks (FCN) such as U-net have been shown to be highly successful at segmentation tasks for medical images analysis and denoising images taken in dark venues. This paper harnesses this well-known deep neural network for the channel decoding challenge recently proven to be suitable for deep neural networks. Previous work have successfully managed to decode convolutional codes using different architectures, such as Recurrent Neural Networks(RNN) and Fully Connected Neural Networks(FCNN) with promising results. However, these approaches are extremely costly in latency, computational resources and memory. This paper shows that taking the approach used in two dimensional image processing, by simple manipulation on the data in the preprocessing phase, achieves better results in a Bit Error Rate(BER) measurement with a large discount on the latency and the number of parameters required to maintain the neural decoder.**

## I. Introduction

The incredible success of deep learning (DL) and neural networks (NNs), previously revolutionized the fields of computer vision and speech processing, has recently triggered further exploration of DL application for communications. This approach was used to derive trainable channel decoders for various codes and channels [1-5], as well as an entire channel-based communication system [6] using the approach of Generative Adversarial Networks (GANs), RF signals and jammers classification [7], and multiple-input multiple-output (MIMO) detectors [8].

DL approach for signal processing for communication often benefits from the so-called sequence-to-sequence NNs, mostly used for natural language processing (NLP) and audio analysis tasks. This analogy should not appear as a surprise, given the sequential nature of the decoding over channel challenge. Thus, one can be easily driven to build a rich variety of different sequence-based recurrent neural network (RNNs) as decoders for a given communication set as presented in Figure 1. It was already been shown [9] that RNNs can improve the performance of channel decoding and code design when truncated back propagation through time concept is used. Most recently, an exhaustive survey over different families of RNN architectures, such as GRUs [10], LSTMs [11] TCNs [12,13], comparing different aspects for decoding convolutional codes was conducted [14].

However, RNNs have major set-backs. These sequence based neural networks are much harder to train, vastly costlier in computational power required and latency time, compared to convolutional neural networks (CNNs). Due to high parallelism capability of modern GPUs, CNNs are much faster to train and are easily used in real time environments.

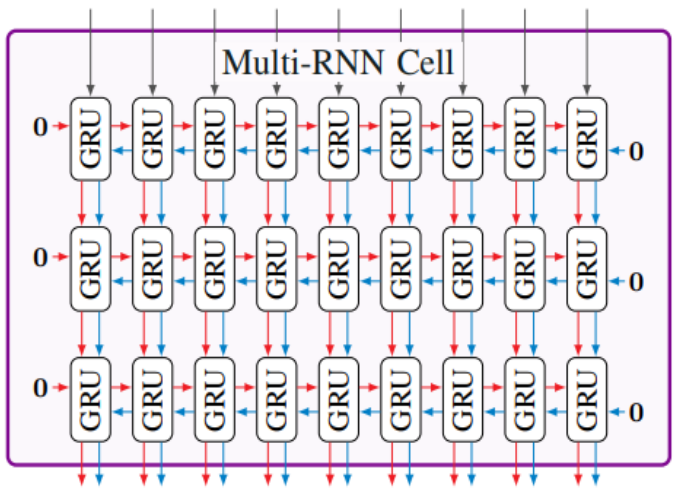

Fig. 1: Multi-layer RNN decoder, taken from [14].

This paper's aim is to introduce CNNs to the channel decoding problem by training a fully convolutional neural network to decode over an additive white Gaussian noise (AWGN) channel. By preordering the data bits in a grid similar to the shape of an image, one can see the almost obvious motivation for this work as driven from the image segmentation challenge usually implemented for medical image processing. In this task, the neural network's goal is to classify each and every pixel. Specifically, in a segmented image, every pixel can be attributed to a predefined class. For binary channel decoding, the number of classes is 2, so that the NN has to attribute every bit to it's original, pre-channel representation. A successful solution to the image segmentation task has been introduced in [15] and has been wildly implemented countless of times, and was also used for denoising of images under extremely dark environments using fully convolutional networks [16], among other different image processing tasks.

For many years, convolutional codes have been used extensively in a variety of communication systems [17-20], benefiting from the famous Viterbi algorithm [21] as a maximum likelihood (ML) decoder as well as relatively simple encoding structure. By providing a clear and optimal baseline for analyzing any neural network performance, convolutional codes provide a helpful benchmark for the neural decoder. It is important to note that the aim of this work is not to surpass and outperform the Viterbi decoder but to provide better solution for decoding convolutional codes via deep learning than the ones previously suggested, thus providing insights into the suitable and efficient NN architectures for signal processing in communications. As claimed in [14], a potential benefit from using NN as decoders can be learning to approximate a low-complex sub-optimal decoder for prohibitively large encoding memories [22]. Furthermore, NN structures can also be significant for practical uses in channel state information (CSI) prediction, equalization and provide input regarding the scalability of auto-encoder systems.

II. Neural Network Architecture

It was already been shown that fully convolutional networks [23, 24] can effectively represent many image processing algorithms [25,26], specifically Unets [15]. The approach implemented in U-net, as depicted in Figure 2, is to supplement a usual contracting network by successive layers, where pooling operators are replaced by upsampling operators. Hence, these layers increase the resolution of the output. In order to localize, high resolution features from the contracting path are combined with the output. A successive convolution layer can then learn to assemble a more precise out based on this information. One important modification in the above architecture is that in the upsampling part there are a large number of feature channels, which allow the network to propagate context information to higher resolution layers. As a consequence, the expansive path is more or less symmetric to the contracting path, and yields a u-shaped architecture. This network does not have any fully connected layers and only uses the valid part of each convolution. In addition, residual connections between the layers are also added to this network.

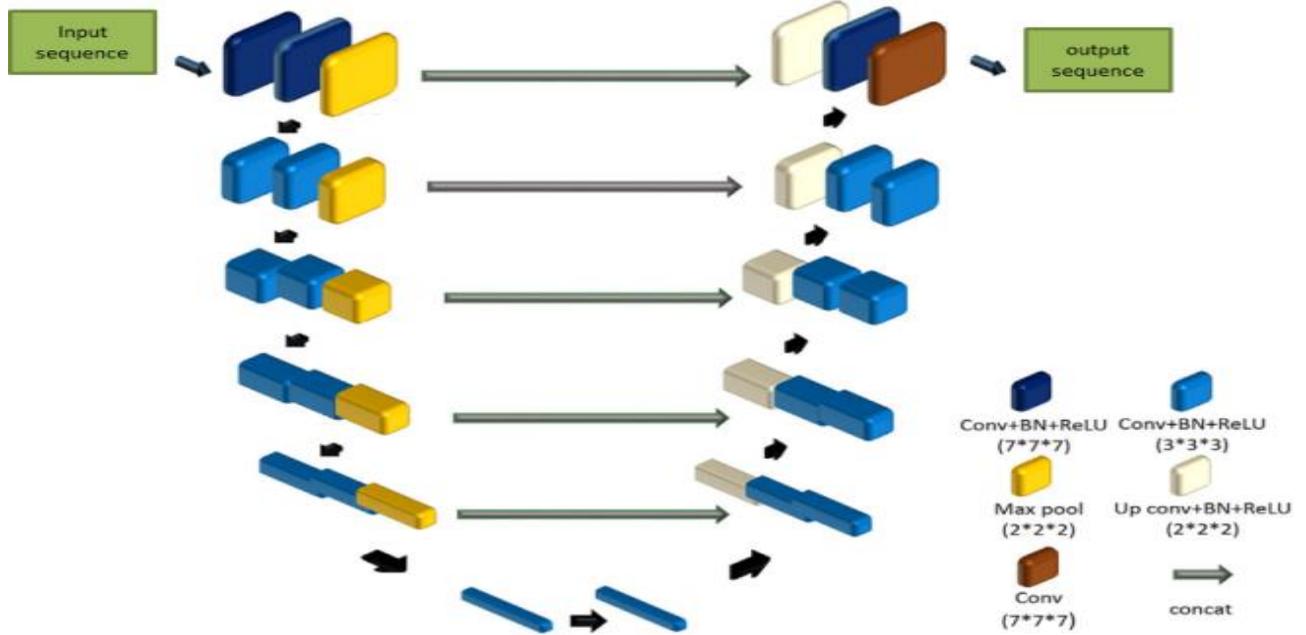

Fig 1: The proposed FCN architecture, U - NET. Taken from [29]

Here, the goal for the neural network system is to decode convolutional codes where every bit of original information is represented by two received sampled bits, as the decoder defined rate is $r = {}^1/_2$. As mentioned above, the input sequence is reshaped to grid after a zero-padding to the original code-word is added, and the output from the network, with the same dimensions, is compared to the input code word for loss calculations.

### III. Deep Learning characterization

1) Optimizer and learning rate: For training the proposed neural network the ADAM[26] optimizer was used. Although there was not much difference observed separating the two for the implemented loss functions, no cumulative research was conducted and is being left for future experiments. It was observed that simply leave the learning rate to be constant as $\eta = 0.001$, throughout this work.

2) Metrics: In order to evaluate the performance of the NN-based decoder, one should obviously consider the BER. In this work the BER is calculated for every mini-batch by hard decision as defined in:

$$BER = \mathbb{E}\left[\frac{1}{l_{1d}}\sum_{k=1}^{l_{1d}} \mathbb{1}_{\{(p_k>0.5)\neq u_k\}}\right] \quad (1)$$

An additional informative metric introduces in [5], is the normalized validation error(NVE). Given an ML decoder exists, one can evaluate a general NN-based decoder be normalizing its BER within a certain SNR range to the optimal achievable BER obtained by the Viterbi decoder

within the same SNR range. The NVE is defined by:

$$NVE(\rho) = \frac{1}{S}\sum_{s=1}^{S} \frac{BER_{NND}(\rho, \rho_{SNR}, s)}{BER_{Viterbi}(\rho_{SNR}, s)} \quad (2)$$

Where S is the number of SNR points, $\rho$ is the design parameter for the neural network pending consideration, and $\rho_{SNR}$ denotes the SNR. This metric provides a straight-forward approach to depict the influence of a certain parameter with respect to the optimal decoder.

3) Loss function: Considering the previous work conducted on sequence base NNs for channel decoding, the cross-entropy loss is a popular choice for the chosen loss function –

$$J_{log} = -\sum_{k=1}^{l_{1d}} u_k \cdot log p_k + (1 - u_k)\log(1 - p_k) \quad (3)$$

When a binary adaptation of (3) is also quite popular option for these kind of comparison.

However, given the motivation for this work is being driven from computer vision, and more specifically image segmentation, in which the loss calculation is based on comparing the original image to the segmented one, it is appropriate to consider using loss function defined in the above mentioned domain.

Image quality evaluation methods can be subdivided into objective and subjective methods. Two well-known objective image quality metrics - given the task in hand one must discard subjective methods - the peak signal to noise ratio (PSNR) and the structural similarity index measure (SSIM) developed by Wang et al. [27]. Given the reference image f and a test image g, both of size MxN, the PSNR between f and g is defined by:

$$PSNR(f, g) = 10\log_{10}(\frac{255^2}{MSE(f,g)}) \quad (4)$$

Where MSE is defined by:

$$MSE(f, g) = \frac{1}{MN}\sum_{i=1}^{M}\sum_{j=1}^{N}(f_{ij} - g_{ij})^2 \quad (5)$$

The PSNR value approaches infinity as the MSE is small, i.e. a higher PSNR value indicates a higher image quality, and a small PSNR implies high numerical differences between the two images. The SSIM is a widely used quality metric considered to be correlated with the quality perception of the human visual perception (HVS). Instead of using traditional error summation methods, the SSIM is designed by modeling any image distortion as a combination of three factors that are loss of correlation luminance distortion and contrast distortion. This metric is defined by:

$$SSIM(f, g) = l(f, g)c(f, g)s(f, g) \quad (6)$$

Where

$$\begin{cases} l(f,g) = \dfrac{2\mu_f\mu_g + C_1}{\mu_f^2 + \mu_g^2 + C_2} \\ c(f,g) = \dfrac{2\sigma_f\sigma_g + C_2}{\sigma_f^2 + \sigma_g^2 + C_2} \\ s(f,g) = \dfrac{\sigma_{fg} + C_3}{\sigma_f\sigma_g + C_3} \end{cases} \quad (7)$$

In the equations above, l(f,g) denotes the luminance comparison function between the two images' mean luminance $\mu_f$ and $\mu_g$, c(f,g) is the contrast comparison function between the standard deviation, $\sigma_f$ and $\sigma_g$, and finally, s(f,g) is the structure comparison function which measures the correlation coefficient when $\sigma_{fg}$ is the covariance for the images. The positive constants $C_1, C_2$ and $C_3$ are used to avoid full denominator.

In [28], it was shown that there is a simple analytical link between the PSNR and the SSIM, however, it was also suggested that the PSNR is more sensitive to additive Gaussian noise than the SSIM.

Keeping the original aim of this paper in mind, it is important to note that even though the notation and terminology in the description above have no relation to communication systems, the metrics taking into account are derived from image processing but borrowed locally for loss function calculation.

### IV. Network Training

The suggested network was trained from scratch, as for now, there were no experiments conducted using transfer learning for decoding communication codes. Throughout the training process, the method suggested in [3] for the SNR considerations was implemented, so that for every batch the SNR is uniformly randomize between 0dB and 8dB. The dataset for training was divided to batch size of 500 samples and contained 150K samples overall. Training proceeds for 500 epochs on GPU NVIDIA RTX 2070 8gb.

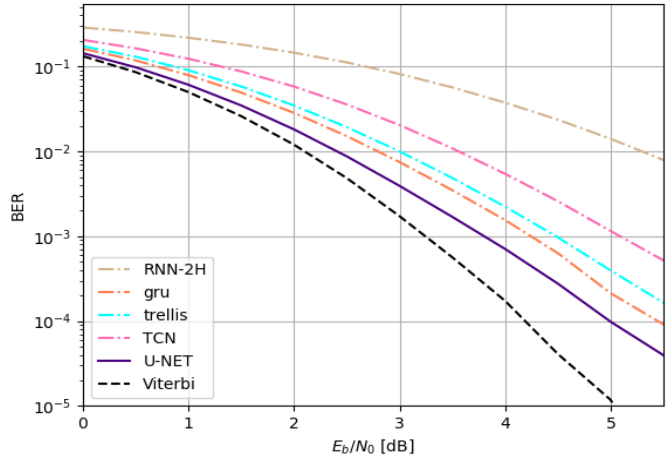

Fig 3: BER performance of referenced NN[14] and the suggested architecture.

### V. Results

In this Section, I will present results which demonstrate that it is possible to use FCN as a convolutional decoder.

#### A. Comparison Of Different NN Designs

Fig 3, shows the BER performance of different NN based decoders. One can notice that the performance of the architecture suggested in this paper is reaching the Viterbi performance and exceeds the performance of the alternative designs.

Fig 4. Shows the BER performance of the proposed design for different convolutional codes, where the chosen codes for reference are the same shown in [14] for easy comparison. One can easily notice that for the codes investigated here the NN-based decoder is able to achieve a performance fairly close to Viterbi performance.

In prior work [14], the NN-based decoder proposed there achieved satisfying results with memory up to $v = 6$, but struggles to do so for codes with $v = 8\ and\ v = 10$. Here one can easily notice that for small memory the decoder is able to generalize solution very well while achieving closer results for higher memory codes.

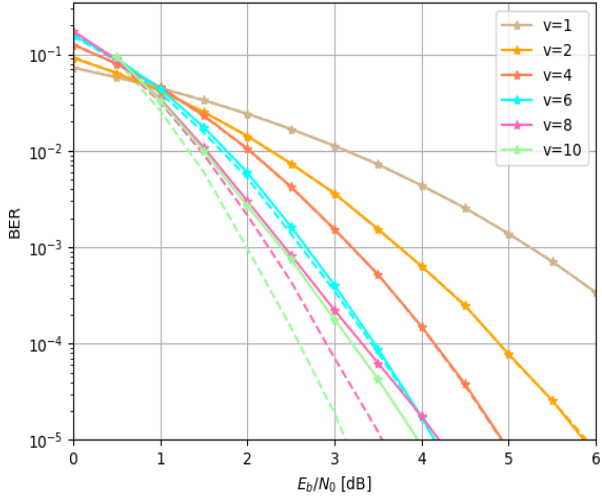

Fig 4: BER performance of the U-net for different codes (dashed lines represents Viterbi).

### B. Latency and parameters

Table 1 presents measurements for the number of parameters and latency of a selected NN based decoder. It is easy to observe that the sequence based NN are heavier and slower than the suggested design in this paper. Furthermore, as expected from the elaborate explanations in the previous chapters in this paper, fully convolutional networks are highly paralleled, shortening the latency time for training and real time use.

### VI. Conclusion

In this paper, it was originally demonstrated that implementing fully convolutional networks as a decoder for communication systems is possible and justified. In addition, a comparison between different popular architectures was presented, from which one can conclude that the u-net is lighter and faster that the alternatives, this without any compensation in the SNR capabilities of the decoder. Further, all three loss function implemented here were able to converge, however, the SSIM function was slightly faster.

| NN | # Parameters | Latency(s) |
|---|---|---|
| TCN | 61 M | 29.73 |
| GRU – 2 | 26 M | 19.24 |
| GRU – 3 | 49 M | 25.6 |
| **U-net** | **13 M** | **9.81** |

Table 1: latency and number of parameters used.

### VII. Future work

Looking forward, this new approach opens new ways to decode convolutional codes using neural networks. Implementing this method for high memory codes, and long code words is a good option for future research, as well as using the transfer learning technique, resume training for specific task or dataset after training from another, to rapidly train the neural net.